\documentstyle[aps,12pt]{revtex}

\begin{document}
\title{Jahn-Teller-effect formation \\
of the non-magnetic ground state for the d$^8$ system*}
\author{Z.Ropka}
\address{Center for Solid State Physics, \'{s}w. Filip 5, 31-150 Krak\'{o}w,}
\author{R.J.Radwanski}
\address{Center for Solid State Physics, \'{s}w. Filip 5, 31-150 Krak\'{o}w;\\
Inst. of Physics, Pedagogical University, 30-084 Krak\'{o}w, Poland.\\
email: sfradwan@cyf-kr.edu.pl}
\maketitle

\begin{abstract}
It is shown that the d$^8$ electron system occurring in the Ni$^{2+}$ ion
can have the non-magnetic ground state in the atomic scale formed by the
Jahn-Teller effect. The tetragonal/trigonal off-cubic lattice distortion for
the octahedral site causes the splitting of the triplet $^3$A$_{2g}$~ into a
singlet-doublet structure with the non-magnetic singlet lying lower provided
the intra-atomic spin-orbit coupling is taken into account.


spin-orbit coupling, Ni$^{2+}$ ion

\end{abstract}


The formation of the non-magnetic ground state of compounds containing Ni$%
^{2+}$~ ions is one of very challenging problem of the 3$d$ magnetism$^{1-4}$%
. Such the ground state is really intriguing because in most of the Ni$^{2+}$%
-ion containing compounds are magnetic (e.g. NiO with Neel temperature of
525 K). Many of these compounds are known as Haldane-gap compounds being
extensively studied after Haldane conjecture$^5$ that one dimensional
Heisenberg magnetic system with the integer spin should possess a
nonmagnetic singlet ground state separated by an energy gap from the excited
magnetic triplet state. Compounds like CsNiCl$_3^1$, organic compounds Ni(C$%
_2$H$_8$N$_2$)$_2$NO$_2$(ClO$_4$) and Ni(C$_2$H$_8$N$_2$)$_2$Ni(CN)$_4$
known as NENP$^{2-3}$ and NENC$^4$ are regarded as being physical
realization of the Haldane system with S=1.

Experimentally, the non-magnetic ground state at lowest temperatures (4.2 K
and lower) is inferred from an anomalous temperature dependence of the
magnetic susceptibility that exhibits a pronounced rounded maximum, e.g. at
55 K for NENP$^{2,3}$ with the vanishing low-temperature susceptibility.
With the increase of temperature this anomalous dependence transforms to the
conventional Curie-Weiss behaviour above, say 100 K, with the effective
moment of 2.5-3.3 ~$\mu _B$. Also a Schottky-like peak in the temperature
dependence of the specific heat is often observed, e.g. at 2.5 K for NENC$^4$%
. These compounds are insulators and contain Ni ions in the divalent state
i.e. Ni$^{2+}$ ions. From crystallographic studies it is known that in many
compounds the nearest ligand surrounding of the Ni cation is well
approximated by the octahedron. A standard crystal-field theory$^{6-8}$
yields for the octahedral site the orbital singlet ground state $^3$A$_{2g}$%
. Forming the orbital singlet, the Ni$^{2+}$ ion in the octahedral site is
not expected to exhibit the Jahn-Teller (J-T) effect.

The aim of the present paper is to show that the non-magnetic state of the $%
d^8$ electronic system can be formed in the atomic scale by the lattice
off-cubic distortion fulfilling the Jahn-Teller theorem provided the
intra-atomic spin-orbit coupling is taken into account. The standard
crystal-field (CEF) theory largely neglects the spin-orbit (s-o) coupling
owing to its weakness in case of 3d ions. Accepting the weakness of the
spin-orbit coupling we have treated the spin-orbit coupling on the same foot
as crystal-field interactions, even of the low symmetry. It turns out that
the d$^8$ system exhibits the Jahn-Teller effect, i.e. the spontaneous
distortion driven by the lowering of the energy by the removal of the
degeneracy of the ground state. It can be theoretically traced provided the
spin-orbit coupling is taken into account. The $d^8$ system is expected to
be realised in Ni$^{2+}$ and Cu$^{3+}$ ions. As we consider the octahedrally
coordinated nickel ion the present discussion, in fact, concerns the
properties of the NiO$_6$ complex in a macroscopic solid. Such the complex
is very often found in realistic systems, where the macroscopic solid is
built from the NiO$_6$ complexes by sharing the wall, the edge or the corner
of the octahedra. For instance, the NaCl structure of the nickel monooxide
NiO is built uo from the edge-sharing NiO$_6$ octahedra.

We assume that eight d electrons of Ni$^{2+}$ and Cu$^{3+}$ ions in the
unfilled 3d shell form the highly-correlated electron system 3$d^8$ as their
orbital and spin movements are strongly correlated within the incomplete
outer shell. According to Hund's rules the $d^8$ system is described by S =1
and L = 3. The resulting 21-fold degeneracy of the lowest term $^3$F is
split by cubic CEF interactions into two orbital triplets $^3$T$_{1,2}$ and
the orbital singlet $^3$A$_2$, the latter being the ground state in case of
the octahedral site (Ref. 6 p.375; Ref. 8 p. 134). The energy positions of
two triplets are usually well above 1.5-2.5 eV (2 eV= 23200K). For further
consideration it is worth noting that the lowest orbital singlet is, in
fact, the triplet in the $\left| LSL_ZS_Z\right\rangle $ space that becomes
physically relevant after involving the spin-orbit coupling. In this space
the triplet can be split by a lattice off-cubic distortion into singlet and
doublet.

The low-energy many-electron structure of the 3$d^8$ system, the splitting
of the $^3$F term and the formation of the singlet non-magnetic ground
state, can be quantitatively traced by studying the single-ion-like
Hamiltonian containing the electron-electron correlations, the crystal
field, the spin-orbit and Zeeman interactions:

\begin{center}
$H_d=H_{el-el}+H_{CF}+H_{s-o}+H_Z(1).$
\end{center}

These interactions have been written in the decreasing-strength succession.
The electron-electron correlations H$_{el-el}$ within the incomplete 3d
shell are taken to be accounted for by the first two Hund's rules. Accepting
them to be realized in the reality we are in the LS space. Then
considerations are confined to the $^3$F term and the Hamiltonian 1 takes
the tractable form

\begin{center}
$H_d^{\prime }=$B$_4(O_4^0+$5$O_4^4)+\lambda L\cdot S+B_2^0O_2^0+\mu _{\text{%
B}}(L+$g$_eS)\cdot B_{ext}(2).$
\end{center}

The first term is the CEF Hamiltonian with the Stevens operators O$_n^m$
that depends on the orbital quantum numbers $L$ and its z-component $L_Z$.
It is separated into the cubic part and an axial term written as $B_2^0O_2^0$
term. The cubic part is usually dominant due to the crystal structure with 6
nearest neighbours forming a distorted octahedron. The second term accounts
for the spin-orbit coupling; $\lambda $ is negative for the Ni$^{2+}$ ion.
The last term accounts for the influence of the magnetic field, the
externally applied in the present case. g$_e$ amounts to 2.0023. The
calculations of the many-electron states of the 3$d^n$ system have been
performed$^9$ by the diagonalization of a (2$L$+ 1)$\cdot $(2$S$+ 1) matrix
associated with the Hamiltonian (2) considered in the $\left|
LSL_ZS_Z\right\rangle $ base. As a result of the diagonalization the
energies and the eigenvectors of the (2$L$+1)$\cdot $(2$S$+1) states are
obtained. The latter ones contain information about the magnetic properties.

The direct calculations of the Hamiltonian (2) have confirmed that the
inclusion of the second-order CEF parameter B$_2^0$, appearing due to an
axial distortion, causes the splitting of the lowest three-fold degenerated $%
^3$A$_{2g}$ state into the singlet and the doublet. Physically, the
singlet-doublet sequence of states depends on the sign of the distortion,
i.e. the elongation or the shrinkage (compression) of the octahedron along
the cube diagonal. The kind of the distortion is related with the sign of
the B$_2^0$ parameter, as shown in Fig. 1.

The performed calculations with the cubic parameter B$_4$= +4.5K, the
spin-orbit coupling $\lambda $=-480K and the tetragonal-distortion parameter
B$_2^0$= +l00K reveal the singlet ground state and the singlet-doublet gap $%
\delta $ of 81 K. The same value of the trigonal-distortion parameter B$_2^0$
causes the gap of 187 K. The parameters used are physically relevant. For
the $d^8$ system placed in the octahedral ligand surrounding B$_4$ is
positive though a bigger value of 20-40 K is expected owing to the $^3$A$%
_{2g}$ - $^3$T$_{2g}$ separation of 1.5-2.5 eV often found for 3d-ion
compounds in optical experiments. This smaller value has been taken here for
the illustration reason associated with the energy scale. The relative
positions of the CEF states are of course preserved.

The existence of the spin-like gap $\delta $ and the (highly-magnetic)
excited doublet state cause interesting temperature dependence of the heat
capacity and the magnetic susceptibility. The heat capacity exhibits a
Schottky-type peak centered at temperature of 0.42$\cdot \delta $ with the
maximal heat of 6.3 J/K mol. The similar maximum is seen in the temperature
dependence of the paramagnetic susceptibility $\chi $ as is presented in
Fig. 2. The parameters used are: B$_4$=+30 K, $\lambda $ = -480 K and B$_2^0$%
=+1000 K. The overall shape of this $\chi (T)$ plot is in very good
agreement with that experimentally observed$^3$ for NENP. The maximum occurs
at 50 K with the value of 18$\cdot 10^{-3}\mu _B/T$ ion. For NENP the
maximum of the susceptibility amounts to 7.2$\cdot $10$^{-3}$ emu/mol that
can be recalculated with the use of the Avogadro number to 13$\cdot
10^{-3}\mu _B/T$ $ion.$ Here is a need for the comment. In the approach
shown above we calculate the contribution to the susceptibility from the one
Ni$^{2+}$ ion placed in the octahedral complex being the part of the
macroscopic solid. The experimental susceptibility is always derived for the
macroscopic system. In the recalculation it is assumed that each Ni atom
contributes equally. This is surely not always the case. Even in the case of
the perovskite structure the orientation of the octahedral MO$_6$ complexes
is not parallel forming a zig-zag structure. It will cause that the
experimentally derived susceptibility is lower. Knowing this fact we
consider the obtained agreement as remarkably good, though we would like to
remind that the exact reproduction of the specific data was not the
intention of this paper.

The main purpose of Fig. 2 is to show the influence of the off-cubic lattice
distortion on the overall temperature dependence of the susceptibility $\chi
(T)$ by comparing with the S=1 behavior. In case of the purely octahedral
crystal field and despite the presence of the s-o coupling $\chi (T)$
follows the Curie law quite well (the curve 2 in Fig. 2) - the $\chi
^{-1}(T) $ plot is practically the straight line yielding the effective
moment of 3.18 $\mu _B.$ This value can be transformed into the
spectroscopic $g$ factor of 2.24 for the effective spin S=1. It means that
our approach allows for the calculations of the g factor - in many
theoretical papers it is simply an assumed parameter basing on experimental
observations (about 2.20 $\mu _B$). It is worth noting that despite of the
non-magnetic ground state and the anomalous low-temperature behavior the
curve 1 mimics the Curie-Weiss behavior at quite low ambient temperatures,
already above, say 150 K.

Finally we would like to point out that the singlet-doublet sequence depends
on the sign of the lattice distortion. The formation of the singlet ground
state is preferred by Nature going spontaneously because then the system
gains energy fulfilling the theorem formulated already in 1937 by Jahn and
Teller. Moreover, the bigger effectiveness of the trigonal distortion in the
realisation of the J-T splitting, visible in Fig. 1, is the predicted effect
that can be experimentally checked. The realisation of the singlet ground
state is also consistent with the Kramers theorem that states that the
singlet ground state can be formed by electrostatic interactions for the
electron system with an even number of electrons. On the other hand we know
a great number of Ni compounds that are magnetically-ordered. Apparently
then, by the formation of the magnetic state the system gains more energy
than by the lattice distortion. The lattice distortion is always eventually
prohibited by the elastic energy.

In conclusion, it has been shown that the non-magnetic singlet ground state
of the highly-correlated $d^8$ electronic systems can be formed on the
atomic scale by the off-cubic lattice distortion provided the spin-orbit
coupling is correctly taking into account. By the realization of the
non-magnetic singlet ground state the NiO$_6$ system fulfils the Jahn-Teller
theorem. The highly-correlated $d^8$ electronic system is thought to be
realised in Ni$^{2+}$ and Cu$^{3+}$ ions even then when they are the part of
a solid. The good reproduction of the experimental data indicates that this
atomic-like structure is largely preserved in the Ni$^{2+}$-ion containing
compounds. The splitting of the lowest orbital singlet by the lattice
distortion via the spin-orbit coupling offers clear physical mechanism
compared e.g. to spin-Hamiltonian considerations where the splitting, known
as a zero-field splitting results from the spin interactions with the
effective spin and to the Haldane-type mechanism where the gap results from
the inter-site spin interactions also with the effective spin and the
neglect of the s-o coupling. For the present calculations the spin-orbit
coupling is essentially important as the splitting of the original orbital
singlet ground state cannot be obtained if the s-o coupling is not taken
into account. Our approach proves the existence of strong correlations
between the local magnetic moment of the $3d$ ion and its local symmetry.

A note added during the referee process. The referees argued that the J-T
theorem does not apply to the system considered in this paper, the ground
state of which is the orbital singlet. A message of this paper is that this
text-book knowledge is not correct - in the 3d$^8$ electronic system
realized in the Ni$^{2+}$ ion the lattice distortion removes the degeneracy
of the ground state orbital singlet that is, in fact, the triplet in the
spin-orbital space [10]. This paper can be also understood as an argument
that the spin-orbital space is physically adequate for description of
electronic and magnetic properties of 3d-ion containing compounds. We add,
that our understanding of the J-T theorem for the d$^8$ system is consistent
with its understanding for the rare-earth systems. The clue of the J-T
effect relies in the fact of the removal of the degeneracy by lattice
distortions with formation of the singlet state, if possible.

Acknowledgements. Authors are indebted to R.Michalski for the help with the
computer preparation of figures and the manuscript.

*The paper has been submitted 14.11.1997 to Phys.Rev.Lett.. An argument of
the referees that the J-T effect is not active in case of the orbital
singlet, in contrary to the present view, has been a reason for the
undertaking by the Managing Editor and the Editor in Chief of
Phys.Rev.Lett., despite of our extensive explanations, a special
discriminating policy. Such the policy is the manipulation of Science by the
Editors of Phys.Rev.Lett. and violates the fundamental scientific rules. Our
request for the publication of our paper with

Fig. 1.) The splitting of the 21-fold degenerated $^3$F term of the 3$d^8$
system. b) The energy level scheme of the 3$d^8$ system under the action of
CEF interactions of the cubic symmetry, produced by octahedrally arranged
ligands, B$_4^0$ 
\mbox{$>$}
0, with the orbital-singlet $^3$A$_{2g}$ ground state. All the orbital
levels have the internal 3-fold spin-degree of freedom, c) the effect of the
s-o coupling according to the present calculations, d and e show the
splitting of the 3-fold degenerated ground state $^3$A$_{2g}$ by the
tetragonal and trigonal distortion. The formation of the singlet ground
state by the given B$_2^0$ parameters should be noticed. The shown splitting
of the lowest orbital-singlet $^3$A$_{2g}$ fulfils the Jahn-Teller theorem
because thanks distortion the system lowers the energy.

Fig. 2. Influence of the lattice off-cubic trigonal distortion (curve 1) on
the temperature dependence of the paramagnetic susceptibility $\chi (T)$ of
the d$^8$ electron system in the octahedral crystal field and in the
presence of the spin-orbit coupling. The parameters used: B$_4$=+30 K, $%
\lambda $ = -480 K and the trigonal distortion B$_2^0$=+1000 K that yield
the electronic structure as shown in the inset, i.e. with non-magnetic
ground state and the magnetic doublet excited state (m=$\pm 2.03$ $\mu _B)$.
The curve 2 shows the susceptibility in the absence of the distortion - it
follows the Curie law though with the effective moment of 3.18 $\mu _B,$
i.e. 12 \% larger than expected for the purely S=1 system.


\begin{references}
\bibitem{9}  The computer program is available on the written request.
\end{references}
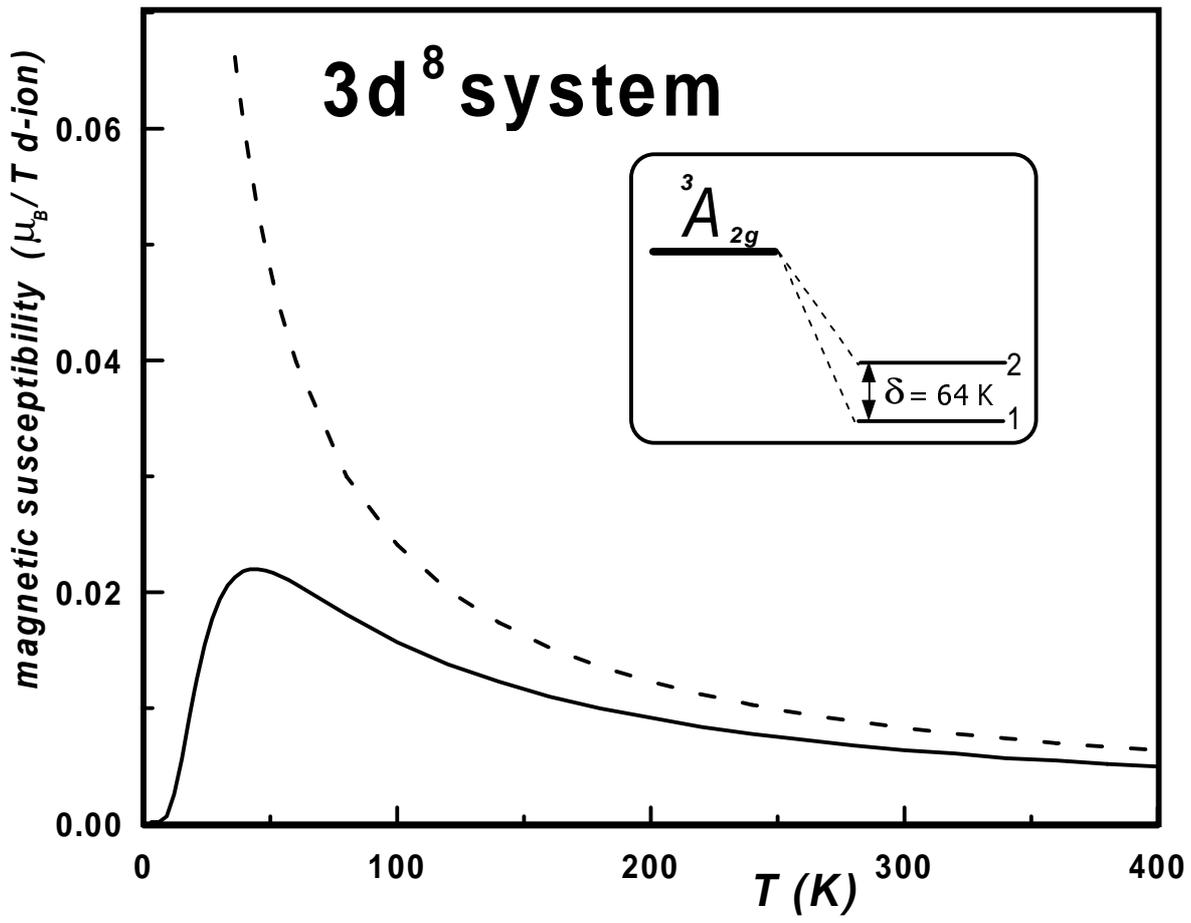
\end{document}